\documentclass[11pt,eqs]{article}
\usepackage{latexsym}

\def\cleq{\setcounter{equation}{0}}

\makeatletter
\newcommand\xleftrightarrow[2][]{%
  \ext@arrow 9999{\longleftrightarrowfill@}{#1}{#2}}
\newcommand\longleftrightarrowfill@{%
  \arrowfill@\leftarrow\relbar\rightarrow}
\makeatother

\title{Open string T-duality in double space
\thanks{Work supported in part by
the Serbian Ministry of Education and Science, under contract No. 171031.}}
\author{B. Sazdovi\'c
\thanks{e-mail: sazdovic@ipb.ac.rs}\\
{\it Institute of Physics,}\\
{\it University of Belgrade,}\\
{\it 11001 Belgrade, P.O.Box 57, Serbia}}

\begin{document}
\maketitle
\begin{abstract}
The role of double space is essential  in new interpretation of T-duality and consequently in an attempt to construct M-theory.
The case of open string is missing in such approach because until now there have been no appropriate formulation of open string T-duality.
In the previous paper \cite{SB1}, we showed how to introduce vector gauge fields $A^N_a$ and $A^D_i$    at the end-points of open string
in order to enable open string invariance under  local gauge transformations of the Kalb-Ramond field and its T-dual
"restricted general coordinate transformations". We demonstrated that gauge fields $A^N_a$ and $A^D_i$ are T-dual to each other.
In the present article we prove that  all above results can be interpreted as coordinate permutations in double space.

\end{abstract}

%%%%%%%%%%%%%%%%%%%%%%%%%%%%%%%%%%%%%%%%%%%%%%%%%%%%%%%%%%%%%%%%%%%%%%%%%%%%%%%%%%%%%
\cleq

\section{Introduction}

It is well known that the M-theory unifies all five consistent  superstring theories by web of T and S dualities.
In order to  formulate the M-theory we should  construct one theory which contain the initial theory
(any of the five consistent) and all corresponding dual ones.

The $2D$ dimensional double space with the coordinates $Z^M= (x^\mu, y_\mu)$ (which components are the
coordinates of initial space $x^\mu$ and its T-dual $y_\mu$)  offers many benefits in interpretation of T-duality.
In fact in such a space, the T-duality transformations can be realized simply by exchanging places of some coordinates $x^a$,  along which we  performed T-duality and the corresponding dual coordinates $y_a$
\cite{SB0,SB}. It contains  the initial and all corresponding T-dual theories.
Realization of such program for T-duality  in the bosonic case has been done: for flat background in Ref.\cite{SB0} and
for the weakly curved background, with  linear  dependence   on coordinates, in Ref.\cite{SB}.
We  hoping that   S-duality, which can be understood as transformation of dilaton background field, can be  successfully  incorporated  in such procedure.

T-duality for superstrings is non-trivial extension of the bosonic case.
In Ref.\cite{NS1} we extended such approach to the type II theories. In fact, doubling all bosonic coordinates we have unified types IIA and IIB theories.
The formulation of M-theory   should  include  T-dualization along fermionic variables, also. T-dualization along all fermionic coordinates in fermionic double space (where we doubled all fermionic variables)
has been considered in Ref.\cite{NS2} .

The remaining step is to extended interpretation of T-duality in double space  (which we earlier propose for the case of the closed string) to the case of the open string, also.
This will be done in the present article.

The difference between open and closed string appears at the open string end points.
Until recently, background fields along Neumann and Dirichlet directions   $A^N_a$ and $A^D_i$
(N and D denote  components with Neumann and Dirichlet boundary conditions)
 are treated in different way \cite{DO,ABB}. The Neumann vector field has been introduced in the Lagrangian trough the coupling with $\dot x^a$.
On the other hand,  Dirichlet vector field has been introduced as a consistency conditions without  contributions to the Lagrangian. In order to realize double space formulation in the open string case we should treat
both Neumann and Dirichlet vector fields in the same way. This has recently been done in Ref.\cite{SB1}.

In Refs.\cite{Zw}  it has been shown how to introduce vector gauge fields  $A^N_a$
in order that   open string  retained symmetries of the closed string.  Note that according to  Ref.\cite{SB1}, beside well known local gauge invariance of the Kalb-Ramond field we used its T-dual "restricted general coordinate transformations", which includes transformations of background fields but not include  transformations of the coordinates.
So, above interpretation of the T-duality in double space will confirm expressions for T-dual  closed string background fields  $G_{\mu \nu}$ and $B_{\mu \nu}$    (as in Refs.\cite{SB0,SB})  and gives the same expressions for T-dual vector fields ${}^\star A_D^a$ and  ${}^\star A_N^i$  as that obtained in Refs.\cite{SB1} with Buscher's procedure.

%%%%%%%%%%%%%%%%%%%%%%%%%%%%%%%%%%%%%%%%%%%%%%%%%%%%%%%%%%%%%%%%%%%%%%%%%%%%%%%%%%%%%
\cleq

\section{T-duality of the open string}

In this section we will introduce some known features of the bosonic string and shortly repeat main results of Ref.\cite{SB1}.
We will adapt T-duality to be in compliance with boundary conditions on the open string end-points.

We will consider vector gauge fields: $A^N_a$ with Neumann boundary conditions  which  compensate not implemented gauge symmetry  of the Kalb-Ramond field  at the open string end-points and
$A^D_i$  with Dirichlet boundary conditions,  which  compensate not implemented restricted  general coordinate transformations at the open string end-points. We will show that field $A^D_i$ is T-dual to the $A^N_a$ one, as well as the general coordinate transformations are T-dual to gauge symmetry.

\subsection{Closed and open Bosonic string}

Let us start with the action for closed bosonic string \cite{Pol,S,Zw}
\begin{equation}\label{eq:action0}
S[x] = \kappa \int_{\Sigma} d^2\xi\sqrt{-g}
\Big[\frac{1}{2}{g}^{\alpha\beta}G_{\mu\nu}[x]
+\frac{\epsilon^{\alpha\beta}}{\sqrt{-g}}B_{\mu\nu}[x]\Big]
\partial_{\alpha}x^{\mu}\partial_{\beta}x^{\nu},
\quad (\varepsilon^{01}=-1) \, .
\end{equation}
It propagates in D-dimensional space-time with background
defined by the space metric $G_{\mu\nu}$ and the Kalb-Ramond field $B_{\mu\nu}$.
We denoted  the string coordinates by $x^{\mu}(\xi),\ \mu=0,1,...,D-1$  and
the intrinsic world-sheet metric by  $g_{\alpha\beta}$. The integration goes over two-dimensional world-sheet $\Sigma$
with coordinates $\xi^\alpha$ ($\xi^{0}=\tau,\ \xi^{1}=\sigma$).

In the conformal gauge $g_{\alpha\beta}=e^{2F}\eta_{\alpha\beta}$ this  action can be rewritten
in terms of   light-cone coordinates
$\xi^{\pm}=\frac{1}{2}(\tau\pm\sigma)$, $\partial_{\pm}= \partial_{\tau}\pm\partial_{\sigma}$ as
\begin{equation}\label{eq:action1}
S  = \kappa \int_{\Sigma} d^2\xi\
\partial_{+}x^{\mu}
\Pi_{+\mu\nu}
\partial_{-}x^{\nu},
\end{equation}
with  following  combination of background fields
\begin{eqnarray}
\Pi_{\pm\mu\nu} =
B_{\mu\nu} \pm\frac{1}{2}G_{\mu\nu}.
\end{eqnarray}

In the string theory, variation  of the action (\ref{eq:action1})  with respect to $x^\mu$ produces not only equation of motion
\begin{equation}\label{eq:eqmot}
\partial_{+}\partial_{-}x^\mu
+\Big{(}\Gamma^{\mu}_{\nu\rho}-B^{\mu}_{\ \nu\rho}\Big{)}
\partial_{+}x^\nu\partial_{-}x^\rho=0  \,  ,
\end{equation}
but also  boundary conditions
\begin{eqnarray}\label{eq:ibc}
\gamma^{(0)}_\mu (x) \delta x^\mu/_{\sigma=\pi} & - & \gamma^{(0)}_\mu  (x) \delta x^\mu/_{\sigma=0} =0 \,       ,
\end{eqnarray}
where $\Gamma^{\mu}_{\nu\rho}$ is Christoffel symbol and   we introduce useful variable
\begin{equation}\label{eq:imp}
\gamma^{(0)}_\mu (x)  \equiv  \frac{\delta S}{\delta {x}^{\prime \mu}}=
\kappa\Big[2B_{\mu\nu} \dot{x}^\nu -G_{\mu\nu}{{x}}^{\prime \nu} \Big]   \,  .
\end{equation}
From now on, we will denote boundary of the open string with $\partial \Sigma$, so that relation (\ref{eq:ibc}) we can  rewrite  as
\begin{eqnarray}\label{eq:ibcs}
\gamma^{(0)}_\mu (x) \delta x^\mu/_{\partial \Sigma} =0 \,       .
\end{eqnarray}

As a consequence of periodicity, the boundary conditions  are trivially satisfied in the closed string case.
In the open string case there are two different ways to satisfy boundary conditions. For some coordinates $x^a$  ($a=0,1, \cdots , p$) we will chose  Neumann boundary conditions, when variations of string end points $\delta x^a/_{\partial \Sigma} \, $  are  arbitrary and for the rest ones
$x^i$ $(i= p+1, \cdots , D-1)$ we will chose  Dirichlet boundary conditions, when edges of the string are fixed $\dot{x}^i/_{\partial \Sigma}=0 \,$.
In order to satisfy Neumann boundary conditions according to (\ref{eq:ibc})  we should take $\gamma^{(0)}_a(x)/_{\partial \Sigma}=0 $.

It s well known   that closed string theory is invariant under following infinitesimal transformations: local gauge transformations of the Kalb-Ramond field
\begin{eqnarray}\label{eq:lgt}
\delta_\Lambda G_{\mu \nu} = 0 \, ,  \qquad  \delta_\Lambda B_{\mu \nu} = \partial_\mu \Lambda_\nu - \partial_\nu \Lambda_\mu \,  ,
\end{eqnarray}
and general coordinate transformations
\begin{eqnarray}\label{eq:gct}
\delta_\xi G_{\mu \nu} = -2 \, (D_\mu \xi_\nu + D_\nu \xi_\mu ) \, ,  \qquad
\delta_\xi B_{\mu \nu} =  -2 \, \xi^\rho B_{\rho \mu \nu} + 2 \partial_\mu (B_{\nu \rho} \xi^\rho) -  2 \partial_\nu (B_{\mu \rho} \xi^\rho)         \,  .
\end{eqnarray}
These transformations are connected by T-duality \cite{DS3,SB1}. Let us stress that according to Ref.\cite{SB1} we are not going to add transformations of the coordinates to (\ref{eq:gct}). So, we will call this
"restricted  general coordinate transformations".

Both above symmetries are failed at the open string end-points. In order to restore these symmetries  the  gauge fields
have to be introduced. To restore local gauge symmetry   we introduce the vector fields $A^N_a$ with Neumann boundary conditions (see Ref.\cite{Zw}), while to restore restricted general coordinate transformations we introduce the vector fields $A^D_i$ with Dirichlet boundary conditions (see Ref.\cite{SB1}). Note that  as a consequence of the boundary conditions only parts of these gauge fields survive.

So, the action for open  bosonic string with above boundary conditions is \cite{SB1}
\begin{eqnarray}\label{eq:giacom}
& S_{open} [x] =   \kappa \int_{\Sigma} d^2\xi   \partial_{_+} x^{\mu} \prod_{+ \mu \nu}     \partial_{_-}x^{\nu}
+  2\kappa \int_{\partial \Sigma} d\tau \Big( A_a^N [x] {\dot x}^a - \frac{1}{\kappa} {A}_i^D [x] G^{-1 i j} \gamma^{(0)}_j (x) \Big)  \nonumber \\
&=   \kappa \int_{\Sigma} d^2\xi   \partial_{_+} x^{\mu} \prod_{+ \mu \nu}     \partial_{_-}x^{\nu}
+ 2\kappa \eta^{\alpha \beta} \int_{\partial \Sigma} d\tau   {\cal A}_{\alpha \mu} [x] \, \partial_\beta x^\mu \,  ,
\end{eqnarray}
where following Ref.\cite{SB1} we introduced  effective variables  ${\cal A}_{\pm \mu} (V) = \{{\cal A}_{\pm a} (V), {\cal A}_{\pm i} (V) \}$ defined as
\begin{eqnarray}\label{eq:calAeq}
 {\cal A}_{\pm a} ( V) \equiv    A_{a}^N ( V)     \,  ,  \qquad     {\cal A}_{\pm i} ( V) =  2 \Pi_{\mp i j}  G^{-1 j k} {A}_k^D ( V)         \,  ,
\end{eqnarray}
and for simplicity we assumed that the metric tensor has a form
\begin{equation}
G_{\mu \nu} = \left (
\begin{array}{cc}
G_{a b}  &  0 \\
0  &   G_{i j}
\end{array}\right )\, .
\end{equation}

We introduced pair of effective vector fields $ {\cal A}_{\alpha \mu} = \{  {\cal A}_{0 \mu} \, ,  {\cal A}_{1 \mu}  \}$ instead of initial one $A_\mu = \{ A_{a}^N \, , A_{i}^D  \}$. So, we doubled the number of
vector fields, but there are two constraints on the effective vector fields
\begin{eqnarray}\label{eq:cvf}
{\cal A}_{1 a} ( V)  =0 \, ,  \qquad     {\cal A}_{0 i} ( V) = - (B G^{-1})_i {}^j  {\cal A}_{1 j} ( V) - {\cal A}_{1 i} (G^{-1} B V)                 \,  .
\end{eqnarray}

In the literature $A_a^N [x]$ is known as massless vector field on Dp-brane while ${A}_i^D [x]$ is known as massless scalar oscillations orthogonal to the Dp-brane.

\subsection{Choice of background}

The space-time  equations of motion are consequence of absence  of the conformal anomaly. For the closed string case
in the lowest order in slope parameter $\alpha^\prime$, it produces \cite{CFMP}
\begin{eqnarray}\label{eq:ste}
&&
\beta_{\mu\nu}^G \equiv   R_{\mu\nu}-\frac{1}{4}B_{\mu\rho\sigma}B_\nu^{\ \rho\sigma}+2D_\mu \partial_\nu\Phi=0,
\nonumber\\
&&
\beta_{\mu\nu}^B \equiv  D_\rho B^\rho_{\ \mu\nu}-2\partial_\rho\Phi B^\rho_{\ \mu\nu}=0,
\nonumber\\
&&
\beta^\Phi \equiv 4(\partial\Phi)^{2}-4D_\mu\partial^\mu\Phi
+\frac{1}{12}B_{\mu\nu\rho}B^{\mu\nu\rho}
+4\pi\kappa(D-26)/3 -R =0        \,  .
\end{eqnarray}
Here
$B_{\mu\nu\rho}=\partial_\mu B_{\nu\rho} +\partial_\nu B_{\rho\mu}+\partial_\rho B_{\mu\nu}$
is the field strength of the Kalb-Ramond  field $B_{\mu \nu}$, and
$R_{\mu \nu}$ and $D_\mu$ are Ricci tensor and
covariant derivative with respect to space-time metric.

With the same reason, for open string case  there are additional space-time equations of motion  \cite{Le}. In our notation they take a form
\begin{eqnarray}\label{eq:osbf}
&  \beta_a \equiv  - \frac{1}{2} {\cal B}_a{}^b \partial_b \Phi + {\cal G}^{-1}_E {}^{b c}  \partial_c  {\cal B}_{b a} +
{\cal G}^{-1}_E {}^{b c} ( \frac{1}{2} {\cal B}_a{}^d B_{d b e} {\cal B}^e {}_c + K^\mu_{a c} B_{\mu \nu} \partial_b f^\nu ) = 0  \,  ,      \nonumber   \\
&  \beta_\mu \equiv  \frac{1}{2} \partial_\mu \Phi   + {\cal G}^{-1}_E {}^{a b} ( \frac{1}{2} {\cal B}_b{}^c B_{\mu a c} -  K_{\mu a b})   = 0       \,  ,
\end{eqnarray}
where
\begin{eqnarray}\label{eq:dcbcg}
 {\cal B}_{a b} = B_{a b}  -2 ( \partial_i A_j^D + \partial_j A_i^D )  \,  ,   \qquad    {\cal G}^E_{a b} = G_{a b } - 4 {\cal B}_{a c}  G^{-1 c d}  {\cal B}_{d b} \,  ,
\end{eqnarray}
and  $K^\mu_{a b}$ is extrinsic curvature.

We will consider the simplest  solutions of the closed string part
\begin{equation}\label{eq:GBP}
G_{\mu\nu}=const \, ,   \quad  B_{\mu\nu}  = const  \, , \quad  \Phi  = const \, , \quad   D=26  \,  ,
\end{equation}
which satisfies  equations (\ref{eq:ste}).   For the open string part  (\ref{eq:osbf}),  we will assume  that vector fields are linear in coordinates \cite{SB1}
\begin{eqnarray}\label{eq:Alc}
A_a^N (x)= A^0_a - \frac{1}{2} F_{a b}^{(a)} x^b \,  , \qquad  {A}_i^D (x) = A^0_i - \frac{1}{4} F_{i j}^{(s)} x^j \,   \,  ,
\end{eqnarray}
so that corresponding field strengths   are constant. The infinitesimal  coefficients  $F_{a b}^{(a)}$ and $F_{i j}^{(s)}$ are defined as
\begin{eqnarray}\label{eq:FaFs}
 F^{(a)}_{a b}= \partial_a A_b^N - \partial_b A_a^N  \, ,     \qquad    F^{(s)}_{i j}= -2 ( \partial_i A_j^D + \partial_j A_i^D ) \, .
\end{eqnarray}
Note that the $ F^{(a)}_{a b}$ is antisymmetric in  $a,b$ indices while the $F^{(s)}_{i j}$ is symmetric in  $i,j$ indices.
Since we are working with plane  Dp-brane the extrinsic curvature is zero and because $\Phi$, ${\cal B}_{a b}$ and $ {\cal G}^E_{a b}$ are
constant, both $\beta_a$  and $\beta_\mu$  vanish.

So, our choice of background fields (\ref{eq:GBP}) and (\ref{eq:Alc}) satisfy all space-time equations of motion.

\subsection{Sigma-model T-duality of the open string}

The T-dualization procedure of the theory described by the action (\ref{eq:giacom}) with background fields (\ref{eq:GBP}) and (\ref{eq:Alc})
has been performed   in Ref.\cite{SB1}.
The T-dualization of the vector background fields $A_a^N$ and ${A}_i^D$  is nontrivial because these fields are coordinate dependent and it is not possible to apply standard Buscher's procedure. Instead,  T-dualization procedure of the Ref.\cite{DS2}, which work in absence of global symmetry, has been applied.

So, applying T-dualization procedure along  all coordinates,   the T-dual action has been obtained
\begin{eqnarray}\label{eq:tdual}
& {}^\star S [y] =   \,
\frac{\kappa^{2}}{2}  \int_\Sigma d^{2}\xi\ \partial_{+}y_\mu \theta_{-}^{\mu\nu} \partial_{-}y_\nu
+2\kappa \int_{\partial \Sigma}  d \tau  \Big( {A}_i^D ( V) G^{-1 i  j}  {\dot y}_j - \frac{1}{\kappa} A_a^N (V) \, {}^\star \gamma_{(0)}^a   \Big)    \nonumber  \\
&= \frac{\kappa^{2}}{2}  \int_\Sigma  d^{2}\xi\ \partial_{+}y_\mu \theta_{-}^{\mu\nu} \partial_{-}y_\nu
+ 2 \kappa \eta^{\alpha \beta}  \int_{\partial \Sigma}  d \tau  \,  {}^\star  {\cal A}^i_\alpha (V) \partial_\beta   y_i   \,  ,
\end{eqnarray}
where
\begin{eqnarray}\label{eq:tpm}
{\theta}^{\mu\nu}_{\pm}&\equiv&
-\frac{2}{\kappa}
(G^{-1}_{E}\Pi_{\pm}G^{-1})^{\mu\nu}=
{\theta}^{\mu\nu}\mp \frac{1}{\kappa}(G_{E}^{-1})^{\mu\nu} \,  ,
\end{eqnarray}
and
\begin{eqnarray}\label{eq:GEt}
G^E_{\mu\nu} & \equiv  G_{\mu\nu}-4(BG^{-1}B)_{\mu\nu}, \qquad
\theta^{\mu\nu} & \equiv  -\frac{2}{\kappa} (G^{-1}_{E}BG^{-1})^{\mu\nu}  \,   ,
\end{eqnarray}
are the symmetric and antisymmetric parts of ${\theta}^{\mu\nu}_{\pm}$. In the literature, $G^E_{\mu\nu}$ is known as open string metric and $\theta^{\mu\nu}$ as non-commutative parameter.

Because T-dual action  (\ref{eq:tdual})  should has the same form as the initial one (\ref{eq:giacom}) but in terms of T-dual fields
we can express  T-dual background fields in terms of initial ones
\begin{eqnarray}\label{eq:tdbfv}
^\star \Pi_{+}^{\mu\nu} = \frac{\kappa}{2} \theta_{-}^{\mu\nu}   \, , \quad   {}^\star {A}^a_D (V) = G_E^{-1 a b}  A_b^N (V)       \, , \quad  {}^\star A^i_N (V) =  G^{-1 i j} {A}_j^D ( V)     \, .
\end{eqnarray}
The first relation can be rewrite as
\begin{equation}\label{eq:tdbf}
^\star G^{\mu\nu} =
(G_{E}^{-1})^{\mu\nu}, \quad
^\star B^{\mu\nu} =
\frac{\kappa}{2}
{\theta}^{\mu\nu}  \, .
\end{equation}

Note that the T-dual vector background fields depend not on $y_\mu$ but on
\begin{eqnarray}\label{eq:solV}
V^\mu = - \kappa \, \theta^{\mu \nu} y_\nu + G^{-1 \mu \nu}_E \, {\tilde y}_\nu   \, ,
\end{eqnarray}
which is function on both $y_\mu$ and  its double
\begin{eqnarray}\label{eq:AtA}
{\tilde y}_\mu= \int (d \tau y^\prime_\mu + d \sigma {\dot y}_\mu)  \, .
\end{eqnarray}

With the help of (\ref{eq:tdbfv}) we can find effective T-dual vector fields in analogy with  relation (\ref{eq:calAeq})
\begin{eqnarray}\label{eq:calAGd}
{}^\star {\cal A}_{\pm }^a (V) = 2 \, {}^\star \Pi_{\mp}^{a b} \, {}^\star  G^{-1}_{b c} \, {}^\star {A}^c_D ( V) = \kappa \, \theta_\pm^{a b} A_b^N  (V)  \, , \qquad
{}^\star  {\cal A}_{\pm }^i ( V) =    {}^\star   A^i_N ( V)  =  G^{-1 i j} A_i^D  (V)   \,  .
\end{eqnarray}

We introduced two effective T-dual vector fields $ {}^\star {\cal A}_\alpha^\mu = \{ {}^\star  {\cal A}_0^\mu \, , {}^\star  {\cal A}_1^\mu  \}$ instead of initial one
${}^\star A^\mu = \{{}^\star  A^{a}_D \, ,{}^\star  A^{i}_N  \}$, but  we have two constraints
\begin{eqnarray}\label{eq:cvfd}
& {}^\star  {\cal A}_{0}^a ( V) =  - 2 ({}^\star B {}^\star G^{-1})^a{}_b  {}^\star {\cal A}_{1}^b ( V)
=  2 ( G^{-1} B)^a{}_b  {}^\star {\cal A}_{1}^b ( V)        \,  ,  \nonumber   \\
&  {}^\star {\cal A}_{1}^i ( V)  = 0   \, .
\end{eqnarray}

For initial and T-dual open strings boundary conditions at the string end-points take a form
\begin{eqnarray}\label{eq:ibc1}
\gamma^{(0)}_\mu \delta x^\mu/_{\partial \Sigma} =  0  \,    , \qquad
{}^\star\gamma_{(0)}^\mu \delta y_\mu/_{\partial \Sigma}  =  0  \,    .
\end{eqnarray}
Here $\gamma^{(0)}_\mu$ (defined in (\ref{eq:imp}) for closed string)  now obtains new infinitesimal  term
\begin{equation}\label{eq:impop}
\gamma^{(0)}_\mu (x)  \equiv  \frac{\delta S}{\delta {x}^{\prime \mu}}=
\kappa\Big[2B_{\mu\nu} \dot{x}^\nu -G_{\mu\nu}{{x}}^{\prime \nu} -2 {\cal A}_{1 \nu} \Delta(\sigma)  \Big]
=  \kappa\Big[2B_{\mu\nu} \dot{x}^\nu -G_{\mu\nu}{{x}}^{\prime \nu} +2  A_i^D  \Delta(\sigma)  \Big]  \,  ,
\end{equation}
while for T-dual theory we have
\begin{eqnarray}\label{eq:dualmom}
{}^\star\gamma_{(0)}^\mu (y)
\equiv \frac{\delta\, ^\star S}{\delta {y}_\mu^\prime}=
\kappa\Big[2  ^\star  B^{\mu\nu} \dot{y}_\nu - {}^\star G^{\mu\nu} {{y}}^\prime_\nu -2 \, \, {}^\star {\cal A}_1^\nu \Delta(\sigma)    \Big]
= \kappa \Big[ \kappa \theta^{\mu\nu} \dot{y}_\nu  -  (G^{-1}_{E})^{\mu\nu} {y}_\nu^\prime +2 \,G^{-1a b}_E  A_b^N \Delta(\sigma) \Big]  \, ,
\end{eqnarray}
where $\Delta(\sigma) \equiv  \delta(\sigma- \pi)  - \delta(\sigma)$.

The  terms with vector field $A_i^D$ in (\ref{eq:impop}) and $A_b^N$  (\ref{eq:dualmom})   are irrelevant in the expressions for actions (\ref{eq:giacom}) and  (\ref{eq:tdual}),
because they appear as infinitesimal of the second order terms.

\subsection{T-duality transformations of the open string}

T-dual transformation lows for the open string, connecting the initial and corresponding T-dual variables take the form \cite{SB1}
\begin{eqnarray}\label{eq:tdtr}
& \partial_{\pm} x^{\mu}  \cong   - \kappa \theta_\pm^{\mu \nu}  \partial_{\pm}y_\nu   \pm 4 \kappa \theta_\pm^{\mu \nu} {\cal A}_{\pm \nu} (V) \Delta(\sigma) \,  , \\ \nonumber
& \partial_{\pm}y_\mu   \cong  -2 \Pi_{\mp \mu \nu} \partial_{\pm} x^{\nu}    \pm  4   {\cal A}_{\pm \mu} (x)  \Delta(\sigma) \,  ,
\end{eqnarray}
where the symbol $\cong$ denotes the T-duality relation.

In fact the second  transformation  (\ref{eq:tdtr})  can be obtained after T-dualization  the T-dual action (\ref{eq:tdual}). The  relations (\ref{eq:tdtr})  are inverse to each other.
Both transformations differ  from the closed string ones by the infinitesimal term which contains vector background fields ${\cal A}_{\pm \mu}$.

In terms of covariant derivatives
\begin{eqnarray}\label{eq:cder}
 D_\pm x^\mu = \partial_\pm x^\mu + 2  (G^{-1})^{\mu \nu} {\cal A}_{\pm \nu} \Delta(\sigma) \,  , \qquad   D_\pm y_\mu = \partial_\pm y_\mu + 2  ({}^\star G^{-1})_{\mu \nu} {}^\star {\cal A}_\pm^{\nu} \Delta(\sigma) \,  ,
\end{eqnarray}
we can rewrite the transformations (\ref{eq:tdtr}) in a simple form
\begin{eqnarray}\label{eq:ctdtr}
 D_{\pm} x^{\mu}  \cong   - \kappa \theta_\pm^{\mu \nu}  D_{\pm}y_\nu    \,  , \qquad
 D_{\pm}y_\mu   \cong  -2 \Pi_{\mp \mu \nu} D_{\pm} x^{\nu}       \,  .
\end{eqnarray}

From first  equation (\ref{eq:tdtr})  we can find the T-dual transformation laws for $\dot{x}^\mu$ and $x^{\prime\mu}$
\begin{eqnarray}\label{eq:dotx}
& \dot{x}^\mu \cong      - \kappa \theta^{\mu \nu} \Big[ {\dot y}_\nu -4 {\cal A}_{1 \nu} \Delta(\sigma)  \Big]   + G_E^{-1}{}^{\mu \nu}
\Big[  y^\prime_\nu -4 {\cal A}_{0 \nu} \Delta(\sigma)  \Big] \nonumber  \\
&= - \kappa \theta^{\mu \nu}  {\dot y}_\nu  + G_E^{-1}{}^{\mu \nu} y^\prime_\nu + 4 \, {}^\star   {\cal A}_1^\mu \Delta(\sigma)
\\\label{eq:xprime1}
& x^{\prime\mu} \cong   - \kappa \theta^{\mu \nu} \Big[  y^\prime_\nu -4 {\cal A}_{0 \nu} \Delta(\sigma)  \Big]   + G_E^{-1}{}^{\mu \nu}
\Big[ {\dot y}_\nu -4 {\cal A}_{1 \nu} \Delta(\sigma)  \Big]  \nonumber  \\
&= - \kappa \theta^{\mu \nu}  y^\prime_\nu  + G_E^{-1}{}^{\mu \nu} {\dot y}_\nu + 4 \,  {}^\star {\cal A}_{0}^\mu \Delta(\sigma)  \Big]
\, ,
\end{eqnarray}
and from the second one  the inverse transformations
\begin{eqnarray}\label{eq:doty}
\dot{y}_\mu &\cong& -2B_{\mu\nu} \dot x^\nu +G_{\mu\nu}x^{\prime\nu} + 4 {\cal A}_{1 \mu} \Delta(\sigma)    \, ,
\\\label{eq:doty1}
y'_\mu &\cong&  G_{\mu\nu}\dot x^\nu-2B_{\mu\nu} x^{\prime\nu} + 4 {\cal A}_{0 \mu} \Delta(\sigma)  \,  .
\end{eqnarray}

Using the expression for the canonical momentum of the original and of the T-dual theory
\begin{equation}\label{eq:mp}
\pi_\mu \equiv \frac{\delta S}{\delta \dot{x}^\mu}=
\kappa\Big[G_{\mu\nu}{\dot{x}}^\nu-2B_{\mu\nu} x^{\prime\nu}  + 2 {\cal A}_{0 \mu} \Delta(\sigma)   \Big] \, ,  \qquad
{}^\star\pi^\mu \equiv \frac{\delta\, ^\star S}{\delta \dot{y}_\mu}= \kappa \Big[   (G^{-1}_{E})^{\mu\nu} \dot{y}_\nu -\kappa \theta^{\mu\nu} y^\prime_\nu  +2\,  {}^\star  {\cal A}_0^\mu \Delta(\sigma)  \Big]  \, ,
\end{equation}
we can rewrite the  transformations  (\ref{eq:xprime1})  and  (\ref{eq:doty1})   in the canonical form
\begin{eqnarray}\label{eq:primex1}
\label{eq:xprime}
\kappa \, x^{\prime\mu} \cong \,^\star\pi^\mu  + 2 \kappa \,  {}^\star  {\cal A}_0^\mu \Delta(\sigma)      \,  ,\qquad \pi_\mu   + 2 \kappa {\cal A}_{0 \mu} \Delta(\sigma) \cong \kappa y^\prime_\mu   \, .
\end{eqnarray}
This relation connect momenta and winding numbers.

We can rewrite the transformations  (\ref{eq:dotx})  and (\ref{eq:doty})    in the form
\begin{eqnarray}\label{eq:dotx1}
\label{eq:xprime}
- \kappa \,    \dot{x}^{\mu}   \cong \, {}^\star \gamma^\mu_{(0)} -  2 \kappa \, {}^\star {\cal A}_1^\mu \Delta(\sigma)  \,  ,\qquad
\gamma_\mu^{(0)} -  2 \kappa {\cal A}_{1 \mu} \Delta(\sigma)  \cong - \kappa \,  \dot{y}_\mu        \, ,
\end{eqnarray}
where $\gamma_\mu^{(0)}$ is defined in  (\ref{eq:impop})  and ${}^\star \gamma^\mu_{(0)}$ in (\ref{eq:dualmom}).

It was shown in Ref.\cite{DS3} that $\pi_\mu$ is generator of general coordinate transformations while $x^{\prime \mu}$ is generator of gauge symmetry. In Ref.\cite{SB1} T-duality relation between
$\dot{x}^{\mu}$ and ${}^\star \gamma^\mu_{(0)}$ (as well as between $\dot{y}_\mu$ and $\gamma_\mu^{(0)}$) has been established.
The  relations (\ref{eq:primex1}) and (\ref{eq:dotx1})  are extension of T-duality to the open string case. The additional ${\cal A}_{\mu}$-dependent  terms stem from variations of the arguments of
vector background fields.

Note that the momentum $\pi_\mu$ and variable $\gamma^{(0)}_\mu (x)$, as well as  $\partial_\alpha x^\mu = \{ {\dot x}^\mu, x^{\prime \mu}  \} $  are components of the same world-sheet vector.
\begin{eqnarray}\label{eq:wsv}
\pi^\alpha_\mu \equiv  \frac{\delta S}{\delta \partial_\alpha x^\mu }  = \{ \pi_\mu, \gamma^{(0)}_\mu (x)  \}           \, .
\end{eqnarray}
From now on we will call $\gamma^{(0)}_\mu (x)$ $\sigma$-momentum. We can rewrite relations (\ref{eq:primex1}) and  (\ref{eq:dotx1}) in the forms
\begin{eqnarray}\label{eq:rewr}
\pi^\alpha_\mu \cong  - \kappa \varepsilon^{\alpha \beta} \partial_\beta y_\mu + 2 \kappa   \eta^{\alpha \beta} {\cal A}_{\beta \mu} \Delta(\sigma)   \, ,  \qquad
{}^\star \pi^{\alpha \mu} \cong  - \kappa \varepsilon^{\alpha \beta} \partial_\beta x^\mu + 2 \kappa   \eta^{\alpha \beta} {}^\star {\cal A}_{\beta}^\mu \Delta(\sigma)    \, .
\end{eqnarray}
Therefore, T-duality interchange Neumann with Dirichlet gauge fields. It also interchange    ${\dot x}^\mu$ and $x^{\prime \mu}$   with ${}^\star \gamma_{(0)}^\mu$ and  ${}^\star \pi^\mu$ as well as
${\dot y}_\mu$ and $y_\mu^\prime$ with $\gamma^{(0)}_\mu$  and $\pi_\mu$.

%%%%%%%%%%%%%%%%%%%%%%%%%%%%%%%%%%%%%%%%%%%%%%%%%%%%%%%%%%%%%%%%%%%%%%%%%%%%%%%%%%%%%
\cleq

\section{ T-dual background  fields of open string in double space}

Following Refs. \cite{Duff,SB0,SB,NS1,NS2} we are going to introduce double space in order to offer simple interpretation of T-dualization as coordinates permutation in double space.
Let us start with T-dual transformation lows (\ref{eq:tdtr}).
We can express them in a useful form,   where on the left hand side we put the terms with world-sheet antisymmetric tensor
$\varepsilon_\alpha{}^\beta$ (note that $\varepsilon_\pm{}^\pm = \pm 1$)
\begin{eqnarray}\label{eq:tdualc}
& \pm \partial_{\pm}y_\mu \cong  G^E_{\mu \nu}  \partial_{\pm}x^\nu
-2 ( B G^{-1})\mu{}^\nu \partial_{\pm}y_\nu  + 8 (\Pi_\pm G^{-1})_\mu{}^\nu   {\cal A}_{\pm \nu} (V)  \Delta(\sigma)    \, ,  \nonumber \\
& \pm \partial_{\pm}x^\mu \cong 2 (G^{-1}  B )^\mu {}_\nu \partial_{\pm}x^\nu +(G^{-1})^{\mu \nu} \partial_{\pm}y_\nu   - 4 G^{-1 \mu \nu}   {\cal A}_{\pm \nu} (x)  \Delta(\sigma)         \, .
\end{eqnarray}

We can rewrite these T-duality relations in the simple form
\begin{equation}\label{eq:tdualds}
\partial_{\pm} Z^M \cong \pm \, \Omega^{MN} \Big[ {\cal{H}}_{NK}  \,\partial_{\pm}Z^K - 2 \Big( {\cal{H}}+  \sigma_3  {\cal{H}} \sigma_3  \Big)_{NK}   A_{\pm}^K (\breve Z_{arg}) \Delta(\sigma)  \Big] \, ,
\end{equation}
where we introduced the double coordinates $Z^M$
and corresponding arguments of background fields $\breve Z_{arg}$
\begin{equation}\label{eq:escoor}
Z^M=\left (
\begin{array}{c}
 x^\mu  \\
y_\mu
\end{array}\right )\, , \qquad
\breve Z_{arg}
=\left |
\begin{array}{c}
 V^\mu   \\
 x^\mu
\end{array}\right |_{D} \, .
\end{equation}

Note a different notation for arguments of background fields, introduced in Ref.\cite{SB}. The double space coordinate $Z^M$ has $2D$ rows, $D$ components of initial coordinates  $x^\mu$  in the upper $D$ rows and
$D$ components of T-dual coordinates $y_\mu$ in the lower $D$ rows.
In  arguments of background fields $\breve Z_{arg} $ in each row there is  the complete $D$
dimensional vector. Rewritten in form of the one column the arguments of background fields are $2 D^2$ dimensional vector.

Because arguments of all background fields in ${\cal A}_{\pm}^M   (\breve Z_{arg})$   and ${}^\star {\cal A}_{\pm}^K  ( {}^\star \breve Z_{arg})$ (see (\ref{eq:tduald}))  are the same in the upper $D$ rows as well as in the lower $D$ rows we can write them in two component notation as in  (\ref{eq:escoor}). We indicated this  with index $D$.

We also introduced  the so called $O(D,D)$ invariant metric $\Omega^{MN}$, the generalized metric ${\cal{H}}_{MN}$ and constant matrix $\sigma_3$
\begin{equation}\label{eq:imgm}
\Omega^{MN}= \left (
\begin{array}{cc}
0 &  1 \\
1  & 0
\end{array}\right )\, ,  \qquad
{\cal{H}}_{MN}  = \left (
\begin{array}{cc}
 G^E_{\mu \nu}  &  -2 \,  B_{\mu\rho}  (G^{-1})^{\rho \nu}  \\
2 (G^{-1})^{\mu \rho} \,  B_{\rho \nu}  & (G^{-1})^{\mu \nu}
\end{array}\right )\, ,
\end{equation}
\begin{equation}
(\sigma_3)_M{}^{N}= \left (
\begin{array}{cc}
1_D &  0 \\
0  & -1_D
\end{array}\right )\, ,
\end{equation}
and  the double gauge fields
\begin{equation}\label{eq:Jpm}
{\cal A}_{\pm}^M   (\breve Z_{arg}) =
\left (
\begin{array}{c}
{}^\star {\cal A}_{\pm}^\mu (V)  \\
{\cal A}_{\pm \mu} (x)
\end{array}\right )   =
\left (
\begin{array}{c}
\kappa \, \theta_\pm^{\mu \nu}  {\cal A}_{\pm \nu} (V)  \\
{\cal A}_{\pm \mu} (x)
\end{array}\right )     \, .
\end{equation}

Note that  as well as in  Refs. \cite{Duff,SB0,SB,NS1,NS2}  all coordinates are doubled. It is easy to check that
\begin{equation}\label{eq:sodd}
{\cal{H}}^T  \Omega {\cal{H}} =\Omega \, ,
\end{equation}
which shows that  manifest $O(D,D)$ symmetry  is automatically incorporated into theory.

\subsection{T-duality in double space along all coordinates}

Let us derive expression for T-dual generalized metric and T-dual double gauge fields following approach of Ref.\cite{SB}.
Then, beside double space coordinate $Z^M$   we should also transform
extended coordinates of the arguments of background fields $\breve Z_{arg}$  (\ref{eq:escoor}).
We will require that the T-duality transformations (\ref{eq:tdualds}) are invariant under transformations
of the  double space coordinates $Z^M$ and $\breve Z_{arg}$
\begin{equation}\label{eq:escoorait}
{}^\star Z^M = {}^\star {\cal T} {}^M {}_N  Z^N  , \qquad
{}^\star  \breve Z_{arg} = {}^\star  \breve {\cal T} \breve Z_{arg} \, .
\end{equation}
We want to offer interpretation for the case where T-dualization has been performed  along all coordinates. So, we are going to exchange all initial with all T-dual coordinates
which is described by  the  matrices  $ {}^\star {\cal T}$ and $ {}^\star \breve {\cal T}$  of the form
\begin{equation}\label{eq:tbt}
 {}^\star {\cal T} = \Omega_2  \otimes 1_D
 =\left (
\begin{array}{cc}
0  &  1_D  \\
1_D  &  0
\end{array}\right ) \,  ,   \qquad
{}^\star \breve {\cal T} =  \Omega_2 \otimes 1_{D^2}
 =\left (
\begin{array}{cc}
0  &  1_{D^2}  \\
1_{D^2}  &  0
\end{array}\right )  \,  .
\end{equation}

The T-dual coordinates ${}^\star Z^M$  and ${}^\star \breve Z_{arg}$  should satisfy  the same relation as initial one  (\ref{eq:tdualds}),  but in terms of T-dual background fields
\begin{equation}\label{eq:tduald}
\partial_{\pm}  {}^\star Z^M \cong \pm \, \Omega^{MN}  \Big[ {}^\star  {\cal{H}}_{NK} \,\partial_{\pm}   {}^\star Z^K - 2 \Big( {}^\star {\cal{H}}+  \sigma_3  {}^\star {\cal{H}} \sigma_3  \Big)_{NK}
{}^\star {\cal A}_{\pm}^K  ( {}^\star \breve Z_{arg}) \Delta(\sigma) \Big] \, .
\end{equation}
This  produces the expression for the dual generalized metric and dual double gauge fields  in terms of the initial ones
\begin{equation}\label{eq:dualgm}
{}^\star  {\cal{H}} \cong  {}^\star  {\cal T} {\cal{H}}   \, {}^\star {\cal T} \,  ,  \qquad  {}^\star {\cal A}_\pm ( {}^\star \breve Z_{arg}) \cong  {}^\star  {\cal T}    {\cal A}_\pm ( \breve Z_{arg})  \, .
\end{equation}
It is well known  \cite{SB0,SB} that the first relation gives the standard T-dual transformations of the metric and Kalb-Ramond fields (\ref{eq:tdbf}). Rewriting the second relation in components,
 with the help of (\ref{eq:tdbf}) and (\ref{eq:Jpm})  we have
\begin{equation}\label{eq:dAA}
 {}^\star {\cal A}_\pm^\mu  \cong   \kappa \, \theta_\pm^{\mu \nu}   {\cal A}_{\pm \nu}  \, .
\end{equation}

Using expressions (\ref{eq:calAeq}) and the first relation   (\ref{eq:tdbfv}) we obtain
\begin{equation}\label{eq:TdDs}
 {}^\star {\cal A}_\pm^a  \cong   \kappa \, \theta_\pm^{a b}   {\cal A}_{\pm b} = 2 \, {}^\star \Pi_\mp^{a b} A_b^N \, , \qquad
  {}^\star {\cal A}_\pm^i  \cong   \kappa \, \theta_\pm^{i j} {\cal A}_{\pm j} = 2 \kappa \, \theta_\pm^{i j}  \Pi_{\mp j k} G^{-1 k q} A_q^D =
  G^{-1 i j} A_j^D  \, .
\end{equation}
On the other hand, the T-dual effective fields  should have the form  (\ref{eq:calAGd})
\begin{equation}\label{eq:TdDsin}
 {}^\star {\cal A}_\pm^a  = 2 \, {}^\star \Pi_\mp^{a b} \, {}^\star G^{-1}_{b c} \, {}^\star    A^c_D  \, , \qquad
  {}^\star {\cal A}_\pm^i  = \, {}^\star   A^i_N  \, .
\end{equation}

From (\ref{eq:TdDs}) and (\ref{eq:TdDsin}) with the help of (\ref{eq:tdbf}) we have
\begin{eqnarray}\label{eq:tdbfvd}
 {}^\star {A}^a_D  = G_E^{-1 a b}  A_b^N        \, , \quad  {}^\star A^i_N  =  G^{-1 i j} {A}_j^D     \, .
\end{eqnarray}
which is just Buscher  T-duality relation  for vector fields (\ref{eq:tdbfv}).

So, inclusion of gauge fields does not change interpretation of T-duality in double space. It is again replacement of the initial and T-dual coordinates which shows that these
transformations are nonphysical.

\subsection{Double space field strength}

If  in addition to  (\ref{eq:escoor}) we  introduce new double fields
\begin{equation}\label{eq:ndf}
{\tilde Z}^M=\left (
\begin{array}{c}
\tilde x^\mu  \\
\tilde y_\mu
\end{array}\right ) \, , \qquad
\partial_M
=\left (
\begin{array}{c}
 \partial_{x \mu}   \\
 \partial^\mu_y
\end{array}\right)  \, , \qquad
{\tilde \partial}_M
=\left (
\begin{array}{c}
 \partial_{{\tilde x} \mu}   \\
 \partial^\mu_{\tilde y}
\end{array}\right)   \, ,
\end{equation}
we can reexpress the field strengths of both initial ant T-dual case  (see Eqs.(5.11) and  (7.31) of Ref.\cite{SB1}) in the form
\begin{eqnarray}\label{eq:FS}
{\cal F}^{M N} =  \Omega^{M K} \Big( {\hat \partial}_{+ K}  \, {\cal A}_+^N  (\breve Z_{arg}) -   {\hat \partial}_{- K} \,   {\cal A}^N (\breve Z_{arg}) \Big) =
\left (
\begin{array}{cc}
{}^\star {\cal F}^{\mu \nu}  &  0  \\
0  &  {\cal F}_{\mu \nu}
\end{array}\right )     \, ,
\end{eqnarray}
where we defined
\begin{eqnarray}\label{eq:pmd}
 {\hat \partial}_{\pm M} =   { \partial}_{M} \pm  {\tilde \partial}_{M}\,  .
\end{eqnarray}

%%%%%%%%%%%%%%%%%%%%%%%%%%%%%%%%%%%%%%%%%%%%%%%%%%%%%%%%%%%%%%%%%%%%%

\cleq

\section{Example: Three torus with $D_1$-brane in double space}

In this section the example of three-torus with $D_1$-brane, considering  in  the Ref.\cite{SB1}, we will present  in double space.
We will show how to perform T-dualization along all coordinates  in double space.

\subsection{Initial theory in double space}

We will start with definition of background fields of the initial theory in double space.
Let us denoted the coordinates of the $D=3$ dimensional torus by $x^{0},x^{1},x^{2}$ and introduce nontrivial components of the  background fields as
\begin{eqnarray}\label{eq:nasapolja}
G_{\mu\nu}=\left(
\begin{array}{ccc}
1 & 0 & 0\\
0 & -1 & 0\\
0 & 0 & -1
\end{array}
\right)  \, , \qquad
B_{\mu\nu} =\left(
\begin{array}{ccc}
0 & B & 0\\
-B & 0 & 0\\
0 & 0 & 0
\end{array}
\right) \, .
\end{eqnarray}

It is easy to find corresponding  effective metric and non-commutativity parameter
\begin{eqnarray}\label{eq:efmncp}
G^E_{\mu\nu}=\left(
\begin{array}{ccc}
G_E & 0 & 0\\
 0 & -G_E & 0\\
0 & 0 & -1
\end{array}
\right)  \, , \qquad
\theta^{\mu\nu} =\left(
\begin{array}{ccc}
0 & \theta & 0\\
- \theta & 0 & 0\\
0 & 0 & 0
\end{array}
\right) \, ,
\end{eqnarray}
where  as we defined in  \cite{SB1}
\begin{eqnarray}\label{eq:GEt}
G_E \equiv  1-4 B^2  \,  , \qquad   \theta \equiv \frac{2 B}{\kappa G_E} \, .
\end{eqnarray}
We will also need expression for combination of background fields
\begin{eqnarray}\label{eq:tpm}
\theta_\pm^{\mu\nu} = \theta^{\mu\nu} \mp \frac{1}{\kappa} G^{-1 \mu \nu}_E  =
\left(
\begin{array}{ccc}
\mp \frac{1}{\kappa G_E} & \theta & 0\\
- \theta & \pm \frac{1}{\kappa G_E}   & 0\\
0 & 0 & \pm \frac{1}{\kappa}
\end{array}
\right) \, .
\end{eqnarray}

According to (\ref{eq:imgm})  it  produces
\begin{eqnarray}\label{eq:calH6}
{\cal H}_{M N}
=  \left(
\begin{array}{cc}
G^E_{\mu \nu}      & -2 (B G^{-1})_\mu{}^\nu \\
2 (G^{-1} B)^\mu{}_\nu  &   (G^{-1})^{\mu \nu}
\end{array}
\right)
=  \left(
\begin{array}{cccccc}
G_E   & 0     & 0 & 0  & 2 B  &   0   \\
0     & - G_E & 0  & 2 B  & 0  &   0   \\
0     & 0 & -1 & 0  & 0  &   0   \\
0     & 2 B & 0 & 1  & 0  &   0   \\
2 B   & 0 & 0 & 0  & -1  &   0   \\
0     & 0 & 0 & 0  & 0  &  -1
\end{array}
\right)  \, .
\end{eqnarray}
Similarly, we have
\begin{eqnarray}\label{eq:calH6s3}
\sigma_3 {\cal H} \sigma_3
=  \left(
\begin{array}{cc}
G^E_{\mu \nu}      & 2 (B G^{-1})_\mu{}^\nu \\
- 2 (G^{-1} B)^\mu{}_\nu  &   (G^{-1})^{\mu \nu}
\end{array}
\right)
=  \left(
\begin{array}{cccccc}
G_E   & 0     & 0 & 0  & -2 B  &   0   \\
0     & - G_E & 0  & - 2 B  & 0  &   0   \\
0     & 0 & -1 & 0  & 0  &   0   \\
0     & - 2 B & 0 & 1  & 0  &   0   \\
- 2 B   & 0 & 0 & 0  & -1  &   0   \\
0     & 0 & 0 & 0  & 0  &  -1
\end{array}
\right)  \, .
\end{eqnarray}

The double space coordinates are
\begin{equation}\label{eq:Zds}
Z^M=\left (
\begin{array}{c}
x^\mu  \\
 y_\mu
\end{array}\right ) =
\left (
\begin{array}{c}
x^0  \\
x^1  \\
x^2  \\
 y_0 \\
 y_1 \\
 y_2
\end{array}\right )  \, , \qquad
\breve Z_{arg} =\left |
\begin{array}{c}
V^\mu  \\
 x^\mu
\end{array}\right |_{D=3} =
\left |
\begin{array}{c}
V^\mu   \\
V^\mu   \\
V^\mu   \\
 x^\mu \\
  x^\mu \\
 x^\mu
\end{array}\right |  \,  ,
\end{equation}
while the double gauge field according to (\ref{eq:Jpm})  takes a form
\begin{equation}\label{eq:Ads}
{\cal A}_\pm^M (\breve Z_{arg}) = \left (
\begin{array}{c}
{}^\star {\cal A}_\pm^\mu  (V)  \\
{\cal A}_{\pm \mu} (x)
\end{array}\right ) =
\left (
\begin{array}{c}
\kappa \, \theta_\pm^{\mu \nu}  {\cal A}_{\pm \nu} (V)  \\
{\cal A}_{\pm \mu} (x)
\end{array}\right )  =
\left (
\begin{array}{c}
{}^\star {\cal A}_\pm^0  (V)   \\
{}^\star {\cal A}_\pm^1  (V)   \\
{}^\star {\cal A}_\pm^2  (V)   \\
{\cal A}_{\pm 0} (x) \\
{\cal A}_{\pm 1} (x) \\
{\cal A}_{\pm 2} (x)
\end{array}\right )  \, .
\end{equation}
Note that dimension of $\breve Z_{arg}$ is $2 \times D^2 = 2 \times 3^2 = 18$.

We will start with  $D_1$-brane define with the Dirichlet boundary conditions  $x^2 (\tau, \sigma)/_{\sigma=0} = x^2 (\tau, \sigma)/_{\sigma=\pi} = const$. It means that we will work with Neumann background fields $A_N^0$ and  $A_N^1$ and Dirichlet background field $A_D^2$ and
according to our convention we will have $p=1$,  $a,b \in \{ 0,1 \}$ and $i,j \in \{2 \}$.

In terms of initial Neumann and  Dirichlet fields we obtain
\begin{equation}\label{eq:Adsi}
{\cal A}_\pm^M (\breve Z_{arg}) =
\left (
\begin{array}{c}
\mp \frac{1}{G_E} {A}_{0}^N (V) + \kappa \theta {A}_{1}^N (V)   \\
 - \kappa \theta {A}_{0}^N (V)  \pm \frac{1}{G_E} {A}_{1}^N (V)   \\
- {A}_2^D  (V)   \\
{A}_{0}^N (x) \\
{A}_{1}^N (x) \\
 \mp {A}_{2}^D (x)
\end{array}\right )  \, ,
\end{equation}
where we used the second  expression  (\ref{eq:efmncp}).

\subsection{T-dual  theory in double space}

On the other hand, for T-dual case we have
\begin{equation}\label{eq:Adsd}
{}^\star {\cal A}_\pm^M (\breve Z_{arg}) =
\left (
\begin{array}{c}
 {\cal A}_{\pm \mu}  (V)  \\
{}^\star {\cal A}_{\pm}^\mu (x)
\end{array}\right ) =
\left (
\begin{array}{c}
2 \Pi_{\mp \mu \nu}\, {}^\star {\cal A}_\pm^\nu  (V)  \\
{}^\star {\cal A}_{\pm}^\mu (x)
\end{array}\right ) =
\left (
\begin{array}{c}
\mp \, {}^\star {\cal A}_\pm^0  (V)  + 2 B \, {}^\star {\cal A}_\pm^1  (V) \\
- 2 B  \, {}^\star {\cal A}_\pm^0  (V) \pm  \, {}^\star {\cal A}_\pm^1  (V)   \\
\pm {}^\star {\cal A}_\pm^2  (V)   \\
{}^\star {\cal A}_{\pm 0} (x) \\
{}^\star {\cal A}_{\pm 1} (x) \\
{}^\star {\cal A}_{\pm 2} (x)
\end{array}\right )  \, ,
\end{equation}
or with the help of (\ref{eq:calAGd}), in terms of T-dual Neumann and  Dirichlet  fields
\begin{equation}\label{eq:Adsdif}
{}^\star {\cal A}_\pm^M (\breve Z_{arg}) =
\left (
\begin{array}{c}
G_E {}^\star {A}^{0}_D (V) \\
- G_E {}^\star {A}^{1}_D (V) \\
\pm {}^\star {A}^{2}_N (V) \\
\mp \, {}^\star {A}^0_D  (x)  - 2 B \, {}^\star {A}^1_D  (x) \\
- 2 B  \, {}^\star {A}^0_D  (x) \mp  \, {}^\star {A}^1_D  (x)   \\
 {}^\star {A}^2_N  (x)
\end{array}\right )  \, .
\end{equation}

Using the second equation (\ref{eq:dualgm}), with the help of (\ref{eq:Adsi}) and (\ref{eq:Adsdif}) we obtain
\begin{equation}\label{eq:Tdeq}
{}^\star {\cal A}_\pm^M ({}^\star \breve Z_{arg}) =
\left (
\begin{array}{c}
G_E {}^\star {A}^{0}_D (x) \\
- G_E {}^\star {A}^{1}_D (x) \\
\pm {}^\star {A}^{2}_N (x) \\
\mp \, {}^\star {A}^0_D  (V)  - 2 B \, {}^\star {A}^1_D  (V) \\
- 2 B  \, {}^\star {A}^0_D  (V) \mp  \, {}^\star {A}^1_D  (V)   \\
 {}^\star {A}^2_N  (V)
\end{array}\right )  \,
\cong  {}^\star  {\cal T}    {\cal A}_\pm ( \breve Z_{arg}) =
\left (
\begin{array}{c}
{A}_{0}^N (x) \\
{A}_{1}^N (x) \\
 \mp {A}_{2}^D (x) \\
\mp \frac{1}{G_E} {A}_{0}^N (V) + \kappa \theta {A}_{1}^N (V)   \\
 - \kappa \theta {A}_{0}^N (V)  \pm \frac{1}{G_E} {A}_{1}^N (V)   \\
- {A}_2^D  (V)
\end{array}\right )  \, ,
\end{equation}
where for this example we have
\begin{equation}\label{eq:Tex}
{}^\star  {\cal T}  =
\left (
\begin{array}{cc}
0 & 1_3 \\
1_3  &  0
\end{array}\right )  \, .
\end{equation}

Note that transition from $ \breve Z_{arg}$ to ${}^\star \breve Z_{arg}$ changes $x^\mu  \leftrightarrow V^\mu$ while operator ${}^\star  {\cal T}$ exchanges first three with last three rows from eq.(\ref{eq:Adsi}).
Expression (\ref{eq:Tdeq})  produces just T-duality relations
\begin{equation}\label{eq:Tdtt}
{}^\star A^0_D =  \frac{1}{G_E} \, A_0^N \, , \qquad  {}^\star A^1_D = -  \frac{1}{G_E} \, A_1^N \, ,  \qquad  {}^\star A^2_N = - A_2^D          \, ,
\end{equation}
in accordance with (\ref{eq:tdbfv}), (\ref{eq:nasapolja}) and (\ref{eq:efmncp}).

The same relation can be obtained with the help of compact notation which produces
${}^\star {\cal A}_\pm^\mu  \cong   \kappa \, \theta_\pm^{\mu \nu}   {\cal A}_{\pm \nu}$,   (see Eq.(\ref{eq:dAA})).
According to (\ref{eq:calAeq}) and  (\ref{eq:calAGd})  we have respectively
\begin{eqnarray}\label{eq:a1}
 {\cal A}_{\pm 0} = A_0^N \, , \quad  {\cal A}_{\pm 1} = A_1^N \, , \quad {\cal A}_{\pm 2} = \mp A_2^D \, ,
\end{eqnarray}
and
\begin{eqnarray}\label{eq:a2}
{}^\star  {\cal A}_{\pm}^0 = \mp \, {}^\star  A^0_D - 2 B \,  {}^\star  A^1_D \, , \,\,
{}^\star  {\cal A}_{\pm}^1 =  - 2 B \,  {}^\star  A^0_D \mp \, {}^\star  A^1_D \, , \,\,
{}^\star  {\cal A}_{\pm}^2 =  {}^\star  A^2_N  .
\end{eqnarray}

Then the equation (\ref{eq:dAA}) takes the form
\begin{equation}\label{eq:Td2}
\left (
\begin{array}{c}
\mp \, {}^\star {A}^0_D    - 2 B \, {}^\star {A}^1_D   \\
- 2 B  \, {}^\star {A}^0_D   \mp  \, {}^\star {A}^1_D     \\
 {}^\star {A}^2_N
\end{array}\right )  \,
 =
 \left(
\begin{array}{ccc}
\mp \frac{1}{ G_E} & \kappa \theta & 0\\
- \kappa  \theta & \pm \frac{1}{ G_E}   & 0\\
0 & 0 & \pm 1
\end{array}
\right)
\left( \begin{array}{c}
 A_0^N    \\
  A_1^N   \\
 \mp A_2^D
\end{array} \right )
= \left (
\begin{array}{c}
\mp \frac{1}{G_E} {A}_{0}^N  + \kappa \theta {A}_{1}^N    \\
 - \kappa \theta {A}_{0}^N   \pm \frac{1}{G_E} {A}_{1}^N    \\
- {A}_2^D
\end{array}\right )  \, ,
\end{equation}
which again produces relation (\ref{eq:Tdtt}).

\subsection{Double space field strength}

The structure  of our example  produces $\gamma^{(0)}_2 = \kappa x^{\prime 2}$ and the action (\ref{eq:giacom}) takes the form
\begin{eqnarray}\label{eq:Sopenpr}
 S_{open} [x] =  \kappa \int_{\Sigma} d^2\xi   \partial_{+}x^{\mu}      \Pi_{+ \mu\nu}  \partial_{-}x^{\nu}
+  2\kappa \int_{\partial \Sigma} d\tau  \Big( A_0^N [x] {\dot x}^0 +  A_1^N [x] {\dot x}^1  + {A}_2^D [x]  x^{\prime 2} \Big)  \,  .
\end{eqnarray}
Note an unusual coupling of Dirichlet part ${A}_2^D$ with $x^{\prime 2}$.

According with (\ref{eq:Alc})  the non trivial vector background fields are
\begin{eqnarray}\label{eq:AlcP}
A_0^N (x)= A^0_0 - \frac{1}{2} F^{(a)} x^1 \,  , \qquad  A_1^N (x)= A^0_1 +  \frac{1}{2} F^{(a)} x^0    \,  , \qquad       {A}_2^D (x) = A^0_2 - \frac{1}{4} F^{(s)} x^2 \,   \,  ,
\end{eqnarray}
where $F^{(a)} \equiv   F_{0 1}^{(a)}= \partial_0 A^N_1 - \partial_1 A^N_0 $ and $F^{(s)} \equiv   F_{2 2}^{(s)} = - 4 \, \partial_2 A^D_2$. Consequently, the field strength of the initial theory is
\begin{eqnarray}\label{eq:fstr}
F_{\mu\nu} = F^{(a)}_{\mu\nu} + \frac{1}{2} F^{(s)}_{\mu\nu}  =
\left(
\begin{array}{ccc}
0  & F^{(a)}  & 0\\
- F^{(a)}  &  0 & 0\\
0 & 0 &  \frac{1}{2} F^{(s)}
\end{array}
\right) \, .
\end{eqnarray}
Note an unusual expression and  unusual  appearance of symmetric field strength $F^{(s)}$.

%%%%%%%%%%%%%%%%%%%%%%%%%%%%%%%%%%%%%%%%%%%%%%%%%%%%%%%%%%%%%%%%%%%%%%%%%%%%%%%%%%%%%
\cleq

\section{ Conclusions}

In the present article we extended interpretation of T-duality in double space to the case of open string.
This includes  consideration of T-duality for the vector gauge fields.

In string theory the  gauge fields appear at boundary of the open string. Their role is to  enable complete local gauge symmetries. In fact, there are two important symmetries of the closed  string theory: local gauge symmetry of the Kalb-Ramond field and general coordinate transformations. In Ref.\cite{SB1} we showed that "restricted general coordinate transformations" (transformations of background fields without transformations of the coordinates)
are T-dual to local gauge symmetry of the Kalb-Ramond field.
Both symmetries are failed at the open string end-points. The function of gauge fields is to restore these symmetries  at the string end-points.

To each symmetry  of the  string theory there is  appropriate gauge field.  As a consequence of the boundary conditions only parts of these gauge fields survive. From gauge field corresponding to local gauge symmetry of the Kalb-Ramond field the components along coordinates with Neumann boundary conditions $A^N_a$ survive. From  gauge field corresponding to restricted general coordinate transformations  the components along coordinates with Dirichlet  boundary conditions $A^D_i$ survive. So, the complete vector field is $A_\mu = \{A^N_a , A^D_i \}$.

In Ref.\cite{SB1} it was shown that known fact that $x^{\prime \mu}$ is T-dual to $\pi_\mu$ produces chain of T-dualities between: restricted general coordinate transformation and local gauge transformations; vector fields with
Neumann  $A^N_a$ and  Dirichlet boundary conditions $A^D_i$.

In the present article we showed that all the above results have simple interpretation in double space.  The double space contains $2D$ coordinates, $D$ initial $x^\mu$ and corresponding $D$ T-dual $y_\mu$. The T-dualization of the present article (along all coordinates) corresponds to the replacement of all initial coordinates $x^\mu$ with all T-dual coordinates $y_\mu$ and all initial arguments of the background fields $x^\mu$ with
all T-dual ones $V^\mu$. Such operation reproduces  all  results  described above.
So, in the open string case complete set of T-duality transformations form the same subgroup of the $2D$ permutation group as in the closed string case.

Let us stress that there is essential difference between our approach and that of Double field theories (DFT) \cite{HZ,HHZ}. In DFT there are two coordinates the initial $x^\mu$ and its double, denoted as $\tilde x_\mu$. The variable  $\tilde x_\mu$ corresponds to our $y_\mu$ but we have additional dual coordinate $\tilde y_\mu$ defined in first relation  (\ref{eq:AtA}).

Consequently, in the double space we are able to represent the backgrounds of all T-dual open string  theories in unified manner as well as in the cases of bosonic \cite{SB0,SB} and type II superstring  theories \cite{NS1}.

This step is  important  ingredient in better understanding  M-theory. We already  explained the role of double space in interpretation of
T-duality and consequently in an attempt to construct M-theory \cite{NS1,NS2}. The present article is extension of these consideration to the case of open string.

%%%%%%%%%%%%%%%%%%%%%%%%%%%%%%%%%%%%%%%%%%%%%%%%%

\end{document}